**Anisotropic superconducting properties of Kagome metal $CsV_3Sb_5$**

Shunli Ni[1,2†], Sheng Ma[1,2†], Yuhang Zhang[1,2†], Jie Yuan[1,2,5], Haitao Yang[1,2,3,5*], Zouyouwei Lu[1,2], Ningning Wang[1,2], Jianping Sun[1,2], Zhen Zhao[1,2], Dong Li[1,2], Shaobo Liu[1,2], Hua Zhang[1,2], Hui Chen[1,2,3,5], Kui Jin[1,2,5], Jinguang Cheng[1,2], Li Yu[1,2,5*], Fang Zhou[1,2,5], Xiaoli Dong[1,2,5*], Jiangping Hu[1,4], Hong-Jun Gao[1,2,3,5], and Zhongxian Zhao[1,2,5]

[1] *Beijing National Laboratory for Condensed Matter Physics and Institute of Physics, Chinese Academy of Sciences, Beijing 100190, China*

[2] *University of Chinese Academy of Sciences, Beijing 100049, China*

[3] *CAS Center for Excellence in Topological Quantum Computation, University of Chinese Academy of Sciences, Beijing 100190, PR China*

[4] *Kavli Institute of Theoretical Sciences, University of Chinese Academy of Sciences, Beijing, 100190, China*

[5] *Songshan Lake Materials Laboratory, Dongguan, Guangdong 523808, China*

[†]These authors contributed equally to this work

[*]E-mail: li.yu@iphy.ac.cn (L.Y.); htyang@iphy.ac.cn (H.Y.); dong@iphy.ac.cn (X.D.)

**Abstract**

We systematically measure the superconducting (SC) and mixed state properties of high-quality $CsV_3Sb_5$ single crystals with $T_c$ ~ 3.5 K. We find that the upper critical field $H_{c2}(T)$ exhibits a large anisotropic ratio of $H_{c2}^{ab}/H_{c2}^{c}$ ~ 9 at zero temperature and fitting its temperature dependence requires a minimum two-band effective model. Moreover, the ratio of the lower critical field, $H_{c1}^{ab}/H_{c1}^{c}$, is also found to be larger than 1, which indicates that the in-plane energy dispersion is strongly renormalized near Fermi energy. Both $H_{c1}(T)$ and SC diamagnetic signal are found to change little initially below $T_c$ ~ 3.5 K and then to increase abruptly upon cooling to a characteristic temperature of ~2.8 K. Furthermore, we identify a two-fold anisotropy of in-plane angular-dependent magnetoresistance in the mixed state. Interestingly, we find that, below the same characteristic $T$ ~ 2.8 K, the orientation of this two-fold anisotropy displays a peculiar twist by an angle of 60° characteristic of the Kagome geometry. Our results suggest an intriguing superconducting state emerging in the complex environment of Kagome lattice, which, at least, is partially driven by electron-electron correlation.

Quasi-two-dimensional (2D) transition-metal Kagome systems serve as an important playground to study various electronic phenomena in the presence of geometric frustration and nontrivial band topology [1-12]. They can host frustrated magnetism, anomalous Hall effect, and various charge orders. The newly discovered $AV_3Sb_5$ ($A$ = K, Rb, Cs) is a new class of Kagome family showing bulk superconductivity at $T_c$ up to 2.5 K [13-15]. Some experimental results suggest unconventional pairing nature in such Kagome system [13,16-19]. In addition, an exotic charge density wave (CDW)-like order can be identified at $T^* \sim 78 - 104$ K in the normal state [11,13-15,18,20-26], and coexists with the superconductivity at lower temperatures. The chirality of such charge order is further verified by scanning tunneling microscopy (STM) measurements [18,20,24,26], which can partially explain the giant anomalous Hall effect (AHE) in the system [23,27]. Recent transport measurements under pressure reveal that a superconducting dome appears as the charge order is suppressed by pressure [17,21,22,28,29]. All these findings suggest the rich and novel physics behind the superconductivity in the $AV_3Sb_5$ system. However, a systematic characterization of the fundamental superconducting (SC) properties such as like lower/upper critical fields ($H_{c1}/H_{c2}$), mixed-state scattering characteristics, magnetic penetration depth ($\lambda$), and coherence length ($\xi$) is still lacking.

In this Letter, we present systematic magnetic and electrical transport measurements of high-quality single crystals of $CsV_3Sb_5$ showing a higher $T_c \sim 3.5$ K than the previous reports. The upper critical field exhibits a large anisotropic ratio $H_{c2}^{ab}/H_{c2}^{c}$ up to 9 at zero temperature, and the fitting to the $H_{c2}(T)$ data requires a minimum two-band effective model. These are consistent with the quasi-2D and multiband nature [23,27] reported previously. Nevertheless, the lower critical field along $ab$-plane ($H_{c1}^{ab}$) is found to be higher than that along $c$-axis ($H_{c1}^{c}$). The in-plane and out-of-plane superconducting coherence length and penetration depth are also deduced. We find that $H_{c1}(T)$ and SC diamagnetic signal change little initially below $T_c \sim 3.5$ K, and then increase abruptly upon cooling to a characteristic temperature ~ 2.8 K. A two-fold anisotropy in the in-plane angular-dependent magnetoresistance (AMR) is identified in the mixed state. Intriguingly, we find that, below the same characteristic $T \sim 2.8$ K, the orientation of this anisotropy displays a twist by an angle of 60° characteristic of the Kagome lattice. Our findings shed new light on the emergence of superconductivity in the presence of the frustrated magnetism and intertwined orders.

The $CsV_3Sb_5$ single crystals were synthesized by the self-flux method [13-15,23,27,29,30]. The x-ray diffraction (XRD) data were collected at room temperature on a diffractometer (Rigaku SmartLab, 9kW) equipped with two Ge (220) monochromators. The magnetic property measurements were conducted down to 1.8 K on a Quantum

Design MPMS-XL1 system, and down to 0.4 K on an MPMS-3 system equipped with an iHe3 insert. The electrical transport properties were measured on a Quantum Design PPMS-9 system under magnetic fields up to 8 T.

The representative XRD pattern of the $CsV_3Sb_5$ single crystals, shown in the upper right inset of Fig. 1a, confirms the single preferred (001) orientation. The lattice parameter $c$ is calculated to be 9.318 Å, consistent with previous reports [23,30]. The double-crystal x-ray rocking curve for the (008) Bragg reflection (Fig. 1a) demonstrates a small crystal mosaic of 0.24° in terms of the full width at half maximum (FWHM), indicating an excellent out-of-plane crystalline perfection. The x-ray $\phi$-scan of the (202) plane in Fig. 1b displays six successive peaks with an equal interval of 60°, in accordance with the hexagonal symmetry of the Kagome lattice. The superconductivity of the $CsV_3Sb_5$ single crystals is confirmed by the superconducting diamagnetism at the onset $T_c \sim 3.5$ K as shown in Fig. 1c, as well as by the zero resistance at ~ 3.0 K in Fig. 1d. We note that the previous work usually reports a superconducting $T_c \sim 2.5$ - 2.8 K [13,17,21,23,24,31] in this Kagome system, which is lower than the $T_c$ and zero-resistance temperature observed here in our sample. The higher $T_c$ of our samples may be due to a sufficient content of the interlayer alkali atoms and almost stoichiometric $V_3Sb_5$ blocks containing the 2D Kagome lattice (see the upper left inset in Fig. 1a). However, the superconducting transition is not as sharp as expected for our high-quality samples. It is noticeable in Fig. 1c that the diamagnetic signal gently sets in below ~ 3.5 Kthen it abruptly drops around ~ 2.8 K, followed by the 100 % superconducting shielding. Such a two-stage-like transition seems common in the $CsV_3Sb_5$ single crystals studied by different groups [13,17,21,23].

In order to obtain the upper critical field $H_{c2}(T)$ of the samples, the data of magnetoresistance $R(T)$ were collected under various fields along the $c$ axis and $ab$ plane. As shown in Fig. 2a and b, the resistive transitions show no significant broadening in the magnetic fields. The temperature dependences of the obtained in-plane $H_{c2}^{ab}(T)$ and out-of-plane $H_{c2}^{c}(T)$ all show positive curvature. Accordingly, the behavior of $H_{c2}(T)$ is well fitted by a two-band model [32] rather than the single-band WHH formula [33] (Fig. 2c), consistent with the multiband nature [23,27]. The zero-temperature $H_{c2}(0)$ is estimated to be 0.8 T for $H//c$ and 7.2 T for $H//ab$. The deduced in-plane coherence length is $\xi_{ab}(0) = 20.3$ nm, and the out-of-plane one $\xi_c(0) = 2.2$ nm. The anisotropy ratio $\gamma = \xi_{ab}/\xi_c \sim 9$, comparable with that of quasi-2D cuprate and iron-based superconductors.

The measurements of lower critical fields $H_{c1}(T)$ can also provide important information of the multiband superconductivity in the Kagome metal $CsV_3Sb_5$. The results of isothermal magnetization $M(H)$ with magnetic field along the $c$ axis and $ab$ plane are presented in Fig. 3a and b, respectively. The temperature dependences of the obtained $H_{c1}{}^{c}$ and $H_{c1}{}^{ab}$ are plotted in Fig. 3c and d, respectively. The lower critical field shows unusual temperature dependence. As the system enters the superconducting state, both $H_{c1}{}^{c}$ and $H_{c1}{}^{ab}$ are nearly independent of temperature, corresponding to the gentle increase of diamagnetic signal with temperature (Fig. 1c). Below the characteristic temperature $T \sim 2.8$ K, both $H_{c1}{}^{c}$ and $H_{c1}{}^{ab}$ rise abruptly with cooling, coinciding with the rapid increase in the Meissner and shielding signal sizes. The temperature dependences of $H_{c1}{}^{c}$ and $H_{c1}{}^{ab}$ are fitted by a semi-classical approach [34] with different models, including single band s-wave, d-wave and a simple two-gap one, as shown respectively in Fig. 3c and d. Obviously, the two-gap model gives the best fit for $H_{c1}{}^{c}(T)$, yielding a London penetration depth $\lambda_{ab}(0) \sim 460$ nm, in agreement with that estimated by tunneling diode oscillator [19].

It is noteworthy that the values of lower critical field $H_{c1}{}^{ab}$ are higher than those of $H_{c1}{}^{c}$, which is reproducibly observed in our samples. This result of $H_{c1}$ suggests that the in-plane effective electron mass ($m_{ab}$) is larger than the out-of-plane one ($m_c$). As suggested in theory [11,12], the electronic physics in this material is likely driven by the Kagome-van Hove points. In this case, the effect of the electron-electron correlation becomes important and the band dispersion near Fermi energy can be strongly renormalized. Thus, combining the presence of van Hove physics and electron-electron correlation, the in-plane effective mass can be strongly enhanced.

To further investigate the unusual electronic and superconducting properties of the $CsV_3Sb_5$ system, the in-plane angular-dependent magnetoresistance was measured under a field of 0.5 T at different temperatures in the mixed state. As shown in Fig. 4a, the AMR exhibits a pronounced *two-fold* rotational symmetry. To our knowledge, such anisotropy in this system has never been reported before. To estimate the strength of relative change of the anisotropic AMR signal, the ratios of $\Delta R/R_{\min} = (R(\theta,T)-R_{\min}(T))/R_{\min}(T)\times 100\%$ are summarized in Fig. 4b by polar-coordinate plots. A large change of $\sim 50\%$ in AMR is observed, despite the quick reduction of the absolute values with lowering temperature, as shown in Fig. 4a. This reflects the emergence of strong anisotropy scatterings under field in the mixed state of the $CsV_3Sb_5$ system.

We note that significant two-fold anisotropy of in-plane AMR has also been observed in a topological superconductor of trigonal $Sr_{0.10}Bi_2Se_3$ under a 0.4 T field in the mixed state [35]. In that case, however, the line-shape of $R(\theta)$ changes significantly with temperature, which is distinct from our observation in $CsV_3Sb_5$ (Fig. 4a). Furthermore, it is intriguing that, as the present hexagonal $CsV_3Sb_5$ is cooled below the characteristic temperature of ~2.8 K, the direction of the maximum AMR rotates by an angle of 60° with respect to the original one, as shown in Fig. 4a and b. Such a peculiar twist of AMR orientation appears to coincide with the Kagome symmetry and its origin is not yet clear. Nevertheless, the anisotropic AMR and its mysterious rotation must relate to the complex electronic and crystallographic environments, which host the emergent anisotropic scatterings intertwined with certain coexisting orders and variable flux dissipation in the superconducting mixed state. The two-fold anisotropy in the mild in-plane field of 0.5 T and its high sensitivity to temperature imply a significant contribution from the spin degree of freedom and a weak coupling to the crystal lattice. Further investigation is certainly required for a better understanding of the physics behind such an exotic rotation.

In summary, we report unusual superconducting properties observed in high-quality single crystals of $CsV_3Sb_5$ with $T_c$ ~ 3.5 K. Our experimental results are consistent with a quasi-2D multiband superconductor. The anisotropic lower critical field suggests a strongly renormalized in-plane effective mass, which indicates the presence of correlated electronic physics due to van Hove points near Fermi energy. We identify two-fold-anisotropic in-plane scatterings under a mild field of 0.5 T in the mixed state, and find a particular twist of 60° of the maximum-scattering direction below a characteristic temperature of ~2.8 K. Moreover, we observe that both $H_{c1}(T)$ and SC diamagnetic signal show concurrent abrupt increases below the same characteristic temperature. These findings shed new light on the emergence of superconductivity in the presence of frustrated magnetism and intertwined orders in the Kagome lattice.

**Acknowledgement**

This work is supported by the National Science Foundation of China (Grant Nos. 11834016, 11888101, 12061131005, 51771224, and 61888102), the National Key Research and Development Projects of China (Grant Nos. 2017YFA0303003 and 2018YFA0305800), the Key Research Program and Strategic Priority Research Program of Frontier Sciences of the Chinese Academy of Sciences (Grant Nos. QYZDY-SSW-SLH001, XDB33010200, and XDB25000000).

**Figure 1**

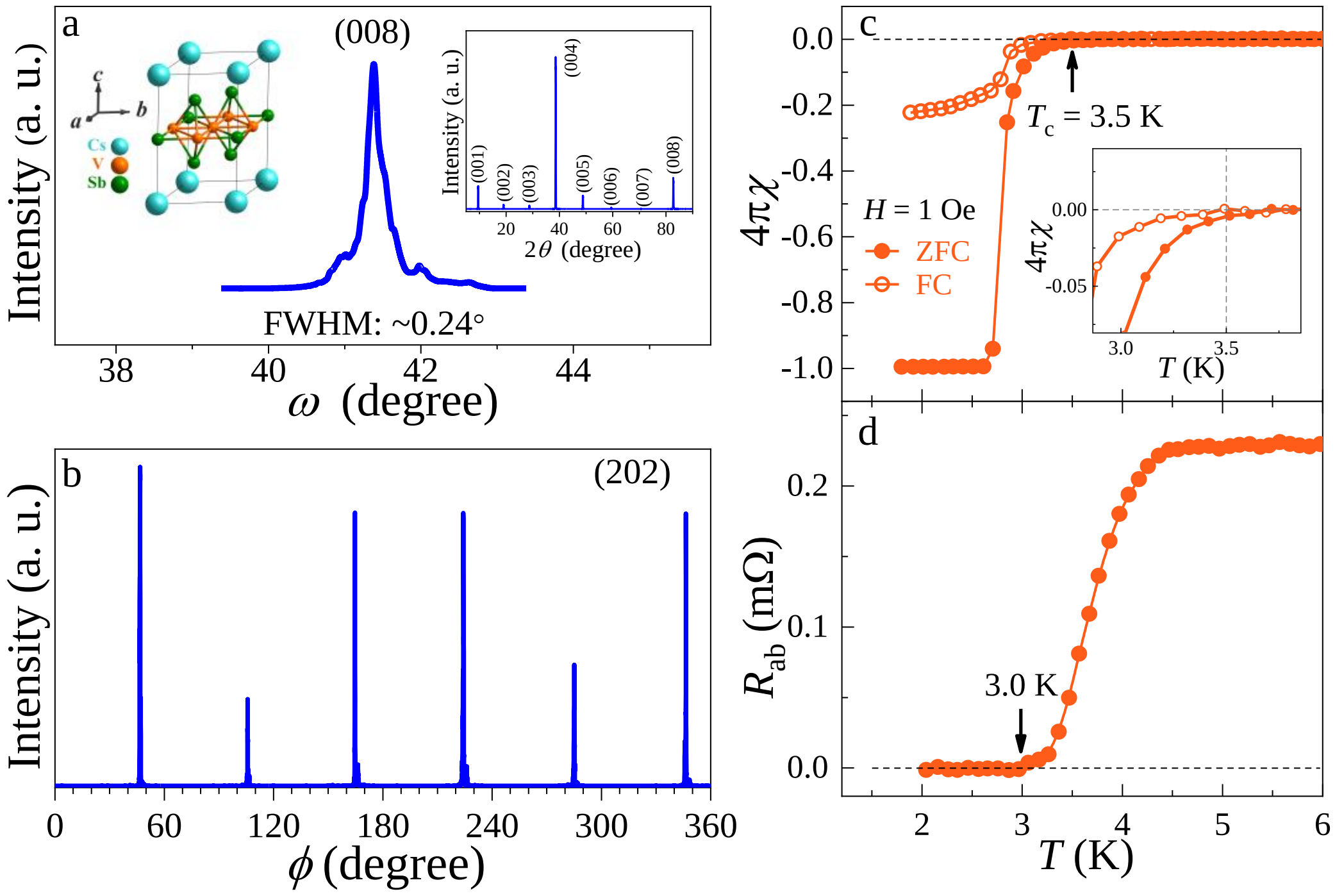

Fig. 1. Crystallographic and superconductivity characterizations of $CsV_3Sb_5$ single crystal. (a) The x-ray rocking curve of (008) reflection shows a small FWHM of 0.24°. The left and right insets show the schematic crystal structure and x-ray $\theta$ - $2\theta$ scan, respectively. (b) The x-ray $\phi$-scan of the (202) plane. (c) The magnetic susceptibilities near the SC transition under zero-field cooling (ZFC) and field cooling (FC) modes. The magnetization data are corrected for the demagnetization factor. Inset : an enlarged view of the superconducting diamagnetic signals. (d) The temperature dependence of the in-plane resistance shows the zero resistance below ~ 3.0 K.

**Figure 2**

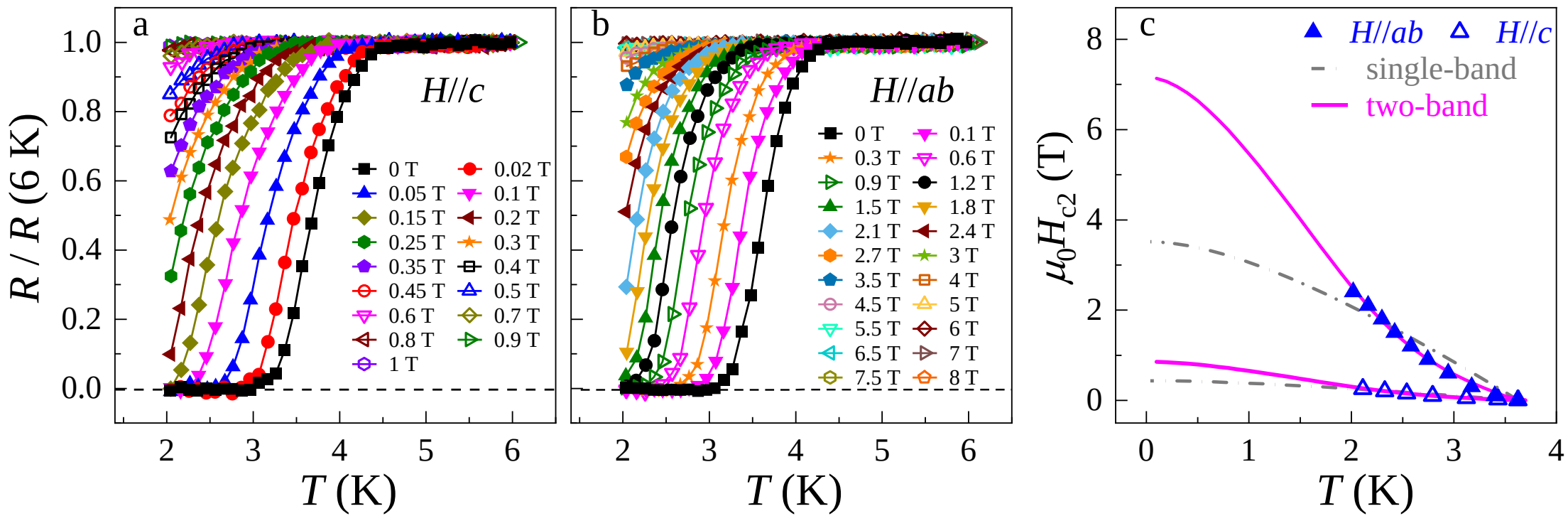

Fig. 2. The normalized resistance under magnetic fields and anisotropic upper critical magnetic fields of the $CsV_3Sb_5$ single crystal. (a), (b) Temperature dependences of normalized in-plane resistance measured under magnetic field along the *c* axis and *ab* plane up to 8T, respectively. (c) Temperature dependences of the upper critical fields, obtained from the *R*(*T*) data (in (a), (b)) at 50 % of the normal-state resistance. The results of fitting by the two-band (pink solid curve) and single-band WHH (gray point-dashed curve) models are presented.

**Figure 3**

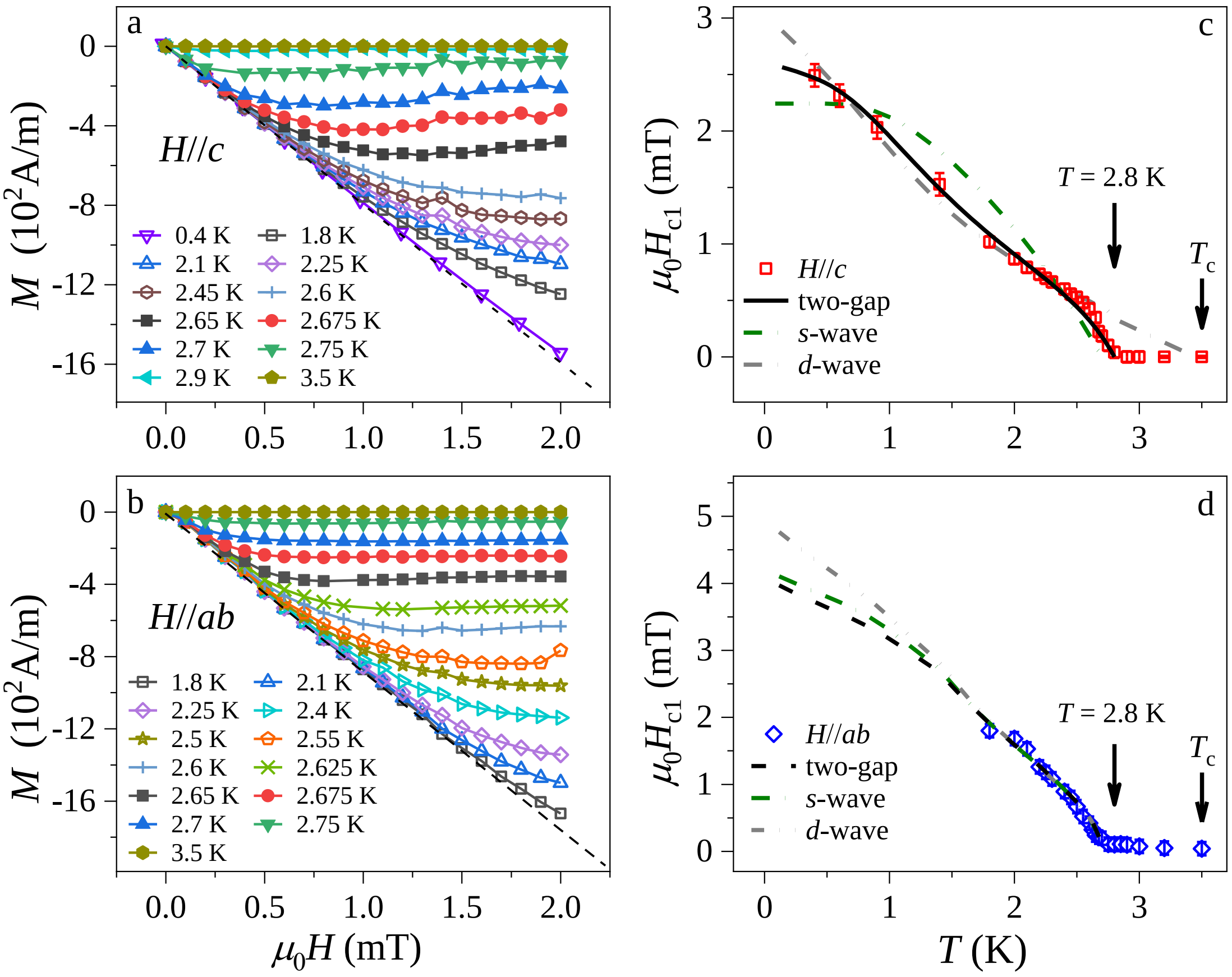

Fig. 3. The isothermal magnetization and anisotropic lower critical field of the $CsV_3Sb_5$ single crystal. (a), (b) The data of isothermal magnetization, corrected for the demagnetization factor, at various temperatures with magnetic field along the *c* axis and *ab* plane, respectively. (c), (d) The temperature dependences of $\mu_0 H_{c1}$ along the *c* axis and *ab* plane, respectively, defined as the fields at which *M*-*H* curves start to deviate from the Meissner line ($M/H = -4\pi$, the dashed lines in (a) and (b)). The results of fitting by the two-gap, s-wave and d-wave models are also presented. The fittings for *H*//*ab* via the three models show no significant difference.

**Figure 4**

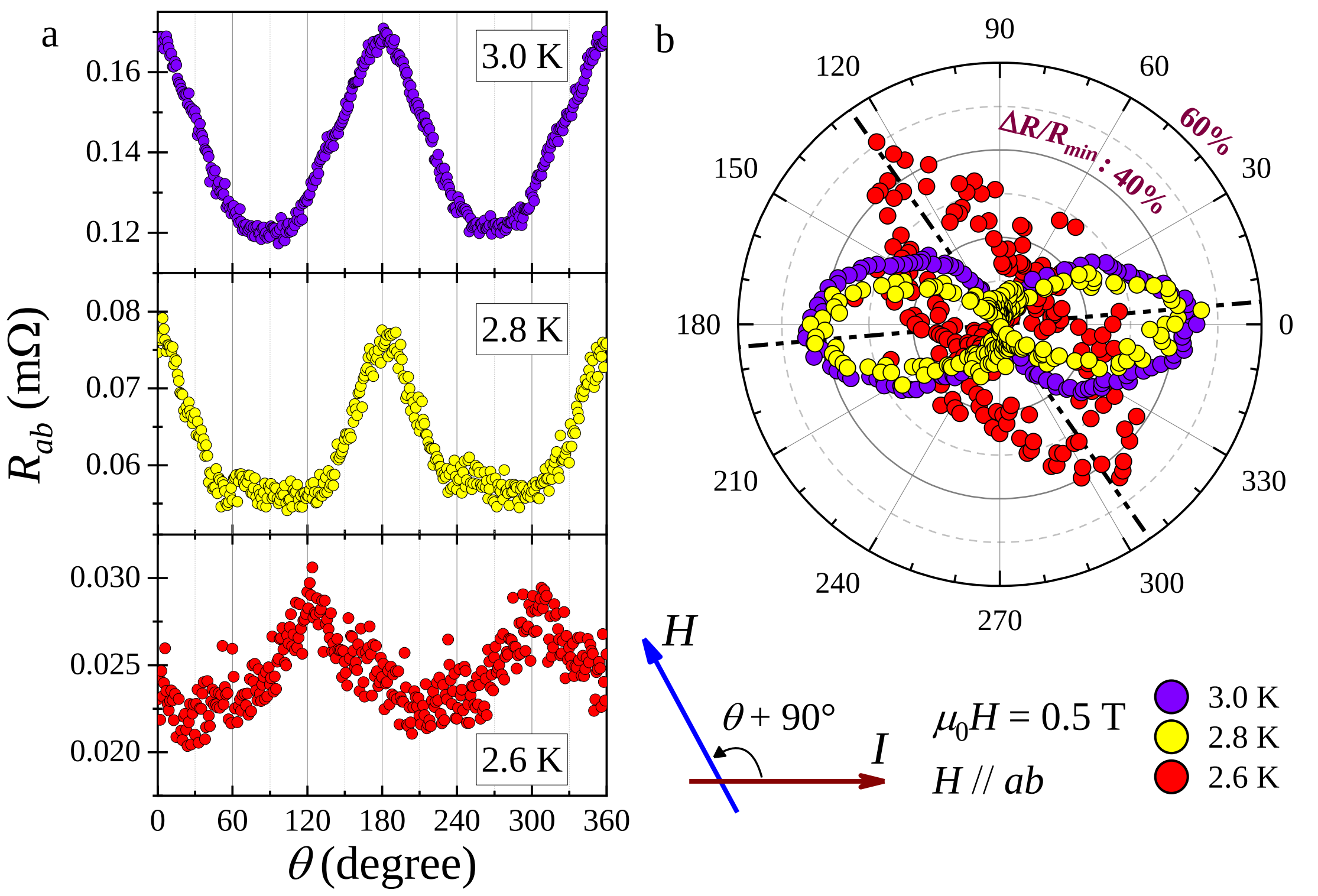

Fig. 4. The in-plane angular-dependent magnetoresistance (AMR) measured in the mixed state. (a) The AMR as a function of $\theta$ at $T$ = 3.0, 2.8 and 2.6K. Here $\theta$ is the angle between the directions of the external field ($H$) and the current ($I$), with $\theta$ = 0° corresponding to $H \perp I$, as illustrated by the schematic diagram. (b) Polar plots of $\Delta R/R_{\min} = (R(\theta,T)-R_{\min}(T))/R_{\min}(T)\times100\%$. The two-fold rotational symmetry in AMR is obvious. The direction of the maximum AMR rotates by an angle of 60° below $T\sim$ 2.8 K.